\documentclass[10pt]{article}
\usepackage{amsfonts}
\usepackage{amsmath}
\usepackage{amssymb}
\usepackage{graphicx}

\begin{document}

\title{An extended Dirac equation in noncommutative space-time}
\author{R. Vilela Mendes\thanks{%
also at Instituto de Plasmas e Fus\~{a}o Nuclear - IST} \thanks{%
rvilela.mendes@gmail.com, rvmendes@fc.ul.pt} \\
Centro de Matem\'{a}tica e Aplica\c{c}\~{o}es Fundamentais, Univ. Lisboa\\
Av. Gama Pinto 2, 1649-003 Lisboa, Portugal}
\date{ }
\maketitle

\begin{abstract}
Stabilizing, by deformation, the algebra of relativistic quantum mechanics a
non-commutative space-time geometry is obtained. The exterior algebra of
this geometry leads to an extended massless Dirac equation which has both a
massless and a large mass solution. The nature of the solutions is
discussed, as well as the effects of coupling the two solutions.
\end{abstract}

\section{The stable Heisenberg-Poincar\'{e} algebra and noncommutative
space-time}

When models are constructed for the natural world, it is reasonable to
expect that only those properties of the models that are robust have a
chance to be observed. Models or theories being approximations to the
natural world, it is unlikely that properties that are too sensitive to
small changes (that is, that depend in a critical manner on particular
values of the parameters) will be well described in the model. If a fine
tuning of the parameters is needed to reproduce some natural phenomenon,
then the model is basically unsound and its other predictions expected to be
unreliable. For this reason it would seem that a good methodological point
of view, in the construction of physical theories, would be to focus on the
robust properties of the models or, equivalently, to consider only models
which are stable, in the sense that they do not change, in a qualitative
manner, when some parameter changes.

This point of view had a large impact in the field of non-linear dynamics,
where it led to the rigorous notion of \textit{structural stability }\cite%
{Andronov} \cite{Smale}. As pointed out by Flato \cite{Flato} and Faddeev 
\cite{Faddeev} the same pattern seems to occur in the fundamental theories
of Nature. In fact, the two physical revolutions of the last century, namely
the passage from non-relativistic to relativistic and from classical to
quantum mechanics are deformations of two unstable Lie algebras to two
stable ones. A mathematical structure is said to be \textit{stable} (or 
\textit{rigid}) for a class of \textit{deformations,} if any deformation in
this class leads to an equivalent (isomorphic) structure. When going from
Galilean to relativistic mechanics, the Galilean algebra, an isolated point,
is deformed to the stable Lorentz algebra and on the transition from
classical to quantum mechanics the unstable Poisson algebra is deformed to
the stable Moyal algebra.

This situation motivated the question of whether the full algebra of
relativistic quantum mechanics, the Heisenberg-Poincar\'{e} algebra, would
itself be stable. The answer was that it is not and that one possible
deformation to a stable one is defined by the following commutators \cite%
{Vilela1}:%
\begin{equation}
\begin{array}{rll}
\lbrack M_{\mu \nu },M_{\rho \sigma }] & = & i(M_{\mu \sigma }\eta _{\nu
\rho }+M_{\nu \rho }\eta _{\mu \sigma }-M_{\nu \sigma }\eta _{\mu \rho
}-M_{\mu \rho }\eta _{\nu \sigma }) \\ 
\lbrack M_{\mu \nu },P_{\lambda }] & = & i(P_{\mu }\eta _{\nu \lambda
}-P_{\nu }\eta _{\mu \lambda }) \\ 
\lbrack M_{\mu \nu },x_{\lambda }] & = & i(x_{\mu }\eta _{\nu \lambda
}-x_{\nu }\eta _{\mu \lambda }) \\ 
\lbrack P_{\mu },P_{\nu }] & = & -i\frac{\epsilon _{4}}{R^{2}}M_{\mu \nu }
\\ 
\lbrack x_{\mu },x_{\nu }] & = & -i\epsilon _{5}\ell ^{2}M_{\mu \nu } \\ 
\lbrack P_{\mu },x_{\nu }] & = & i\eta _{\mu \nu }\Im  \\ 
\lbrack P_{\mu },\Im ] & = & -i\frac{\epsilon _{4}}{R^{2}}x_{\mu } \\ 
\lbrack x_{\mu },\Im ] & = & i\epsilon _{5}\ell ^{2}P_{\mu }%
\end{array}
\label{1.1}
\end{equation}%
In this algebra, that will be denoted $\Re _{\ell ,R}=\{M_{\mu \nu },P_{\mu
},x_{\mu },\Im \}$, $M_{\mu \nu }$ are the generators of the Lorentz group, $%
P_{\mu }$ and $x_{\mu }$ the momenta and coordinates, $\Im $ a non-trivial
operator that replaces the center of the Heisenberg algebra and $\epsilon
_{4},\epsilon _{5}$ are $\pm $ signs. Velocities and actions are in units of 
$c$ and $\hbar $ (that is $c=\hbar =1$). The stable algebra $\Re _{\ell ,R}$
is isomorphic to the algebra of the $6-$dimensional pseudo-orthogonal group
with metric 
\begin{equation}
\eta _{aa}=(1,-1,-1,-1,\epsilon _{4},\epsilon _{5}),\bigskip \ \epsilon
_{4},\epsilon _{5}=\pm 1  \label{1.2}
\end{equation}%
The nonvanishing right hand side of the $[P_{\mu },P_{\nu }]$ commutator
simply means that flat space is an isolated point in the set of arbitrarily
curved spaces. This is why Faddeev \cite{Faddeev} points out that Einstein's
theory of gravity may also be considered as a deformation in a stable
direction. Einstein's theory is based on curved pseudo Riemann manifolds.
Therefore, in the set of Riemann spaces, Minkowski space is a kind of
degeneracy whereas a generic Riemann manifold is stable in the sense that in
its neighborhood all spaces are curved. However, as long as one is concerned
with the kinematical group of the tangent space to the space-time manifold,
and not with the group of motions in the manifold itself, it is perfectly
consistent to take $R\rightarrow \infty $ and this deformation would be
removed. In particular, because the curvature is not a constant, $R$ cannot
have the status of a fundamental constant.

In contrast, for the other stabilizing deformation, associated to the $\ell $
constant, there is no obvious reason to remove it and $\ell $ might play the
role of a new fundamental constant. Taking the $\Re _{\ell ,\infty }$
algebra as the kinematical algebra of tangent space, the main features are
the non-commutativity of the $x_{\mu }$ coordinates and the fact that $\Im $%
, previously a trivial center of the Heisenberg algebra, becomes now a
non-trivial operator. These are however the minimal changes that seem to be
required if stability of the algebra of observables (in the tangent space)
is a good guiding principle. Two constants define this deformation. One is $%
\ell $, a \textit{fundamental length}, the other the sign $\epsilon _{5}$.

The algebra $\Re _{\ell ,\infty }$ is seen to be the algebra of the
pseudo-Euclidean groups $E(1,4)$ or $E(2,3)$, depending on whether $\epsilon
_{5}$ is $-1$ or $+1$. It has a simple representation by differential
operators in a five-dimensional space with coordinates $(\xi _{0},\xi
_{1},\xi _{2},\xi _{3},\xi _{4})$%
\begin{equation}
\begin{array}{rll}
P_{\mu } & = & i\frac{\partial }{\partial \xi ^{\mu }}+iD_{P_{\mu }} \\ 
M_{\mu \nu } & = & i(\xi _{\mu }\frac{\partial }{\partial \xi ^{\nu }}-\xi
_{\nu }\frac{\partial }{\partial \xi ^{\mu }})+\Sigma _{\mu \nu } \\ 
x_{\mu } & = & \xi _{\mu }+i\ell (\xi _{\mu }\frac{\partial }{\partial \xi
^{4}}-\epsilon _{5}\xi ^{4}\frac{\partial }{\partial \xi ^{\mu }})+\ell
\Sigma _{\mu 4} \\ 
\Im  & = & 1+i\ell \frac{\partial }{\partial \xi ^{4}}+i\ell D_{\xi ^{4}}%
\end{array}
\label{1.3}
\end{equation}%
The set ($\Sigma _{\mu \nu },\Sigma _{\mu 4}$) is an internal spin operator
for the groups $O(4,1)$ (if $\epsilon _{5}=-1$) or $O(3,2)$ (if $\epsilon
_{5}=+1$) and $D_{P_{\mu }}$ and $D_{\xi ^{4}}$ are derivations operating in
the space where ($\Sigma _{\mu \nu },\Sigma _{\mu 4}$) acts. For practical
calculations, in particular for the construction of quantum fields, it may
be convenient to use this representation. Notice however that only the
Poincar\'{e} part of $E(1,4)$ or $E(2,3)$ corresponds to symmetry operations
and only this part has to be implemented by unitary operators. Also,
although an extra dimension is used in the representation space, the
space-time coordinates are still only four, noncommutative ones. Physical
consequences of the non-commutative space-time structure implied by the $\Re
_{\ell ,\infty }$ algebra have already been explored in a series of
publications \cite{Vilela2}-\cite{Vilela8}. Depending on the sign of $%
\epsilon _{5}$ the time ($\epsilon _{5}=+1$) or one space variable ($%
\epsilon _{5}=-1$) will have discrete spectrum. In any case $\ell $, a new
fundamental constant, sets a natural scale for time and length. If $\ell $
is of the order of Planck's length, observation of most of the effects
worked out in the cited references will be beyond present experimental
capabilities. However, if $\ell $ is much larger than Planck%
\'{}%
s length (for example of order $10^{-27}-10^{-26}$ seconds) the effects
might already be observable in the laboratory or in astrophysical
observations. Some of the most noteworthy effects arise from the
modification of the phase space volume and from interference effects.

However, most of the consequences worked out in the references \cite{Vilela2}%
-\cite{Vilela6} are rather conservative, in the sense that they simply
explore the nonvanishing of the right-hand-side of the commutators of
previously commuting variables. Deeper consequences are to be expected from
the radical change from a commutative to a non-commutative space-time
geometry, in particular from the new differential algebra associated to the
geometry. One such consequence will be described in this paper. The new
geometry was studied in Ref.\cite{Vilela7} to which I will refer for details
and notation.

\section{The differential algebra and an extended Dirac equation}

In the framework of the non-commutative geometry implied by the deformed
algebra, the differential algebra may be constructed either by duality from
the derivations of the algebra or from the triple $\left( H,\pi (U_{\Re
}),D\right) $, where $U_{\Re }$ is the enveloping algebra of $\Re _{\ell
,\infty }$, to which a unit and, for later convenience, the inverse of $\Im $%
, are added. 
\begin{equation}
U_{\Re }=\{x_{\mu },M_{\mu \nu },p_{\mu },\Im ,\Im ^{-1},1\}  \label{3.13}
\end{equation}%
$\pi (U_{\Re })$ is a representation of the $U_{\Re }$ algebra in the
Hilbert space $H$ and $D$ is the Dirac operator, the commutator with the
Dirac operator being used to generate the one-forms. In a general
non-commutative framework \cite{Connes2} \cite{Connes-Marcolli} it is not
always possible to use the derivations of the algebra to construct by
duality the differential forms. In particular, many algebras have no
derivations at all. However when the algebra has enough derivations it is
useful to consider them \cite{Dubois1} \cite{Dubois2} because the
correspondence of the non-commutative geometry notions to the classical ones
becomes very clear. One considers here the set $V$ of derivations with basis 
$\{\partial _{\mu },\partial _{4}\}$ defined as follows\footnote{%
Notice that the definition of $\partial _{4}$ here is slightly different
from the one in Ref.\cite{Vilela7}.}%
\begin{equation}
\begin{array}{lll}
\partial _{\mu }(x_{\nu }) & = & \eta _{\mu \nu }\Im \\ 
\partial _{4}(x_{\mu }) & = & -\epsilon _{5}\ell p_{\mu }\Im \\ 
\partial _{\sigma }(M_{\mu \nu }) & = & \eta _{\sigma \mu }p_{\nu }-\eta
_{\sigma \nu }p_{\mu } \\ 
\partial _{\mu }(p_{\nu }) & = & \partial _{\mu }(\Im )=\partial _{\mu }(1)=0
\\ 
\partial _{4}(M_{\mu \nu }) & = & \partial _{4}(p_{\mu })=\partial _{4}(\Im
)=\partial _{4}(1)=0%
\end{array}
\label{3.14}
\end{equation}%
In the commutative ($\ell =0$) case a basis for $1$-forms is obtained, by
duality, from the set $\{\partial _{\mu }\}$. In the $\ell \neq 0$ case the
set of derivations $\{\partial _{\mu },\partial _{4}\}$ is the minimal set
that contains the usual $\partial _{\mu }$'s, is maximal abelian and is
action closed on the coordinate operators, in the sense that the action of $%
\partial _{\mu }$ on $x_{\nu }$ leads to the operator $\Im $ associated to $%
\partial _{4}$ and conversely.

The operators that are associated to the physical coordinates are just the
four $x_{\mu }$, $\mu \in (0,1,2,3)$. However, an additional degree of
freedom appears in the set of derivations. This is not a conjectured extra
dimension but simply a mathematical consequence of the algebraic structure
of $\Re _{\ell ,\infty }$ which, in turn, was a consequence of the
stabilizing deformation of relativistic quantum mechanics. No extra
dimension appears in the set of physical coordinates, because it does not
correspond to any operator in $\Re _{\ell ,\infty }$. However the
derivations in $V$ introduce, by duality, an additional degree of freedom in
the exterior algebra. For example, all quantum fields that are Lie
algebra-valued connections will pick up additional components. These
additional components, in quantum fields that are connections, are a
consequence of the length parameter $\ell $ which does not depend on its
magnitude, but only on $\ell $ being $\neq 0$.

The Dirac operator \cite{Vilela7} is%
\begin{equation}
D=i\gamma ^{a}\partial _{a}  \label{3.15}
\end{equation}%
with $\partial _{a}=\left( \partial _{\mu },\partial _{4}\right) $, the $%
\gamma $'s being a basis for the Clifford algebras $C\left( 3,2\right) $ or $%
C\left( 4,1\right) $%
\begin{equation}
\begin{array}{ccc}
\left( \gamma ^{0},\gamma ^{1},\gamma ^{2},\gamma ^{3},\gamma ^{4}=\gamma
^{5}\right)  &  & \epsilon _{5}=+1 \\ 
\left( \gamma ^{0},\gamma ^{1},\gamma ^{2},\gamma ^{3},\gamma ^{4}=i\gamma
^{5}\right)  &  & \epsilon _{5}=-1%
\end{array}
\label{3.16}
\end{equation}%
How to construct quantum, scalar, spinor and gauge fields, as operators in $%
U_{\Re }$, has been described in \cite{Vilela7}. In particular the role of
the additional dimension in the exterior algebra, on gauge interactions, has
been emphasized (see also \cite{Vilela6}). Here another potentially
interesting consequence for spinor fields will be described. Because 
\begin{equation}
\left[ p_{\mu },e^{-\frac{i}{2}k_{\nu }\left\{ x^{\nu },\Im ^{-1}\right\}
_{+}}\right] =k_{\mu }e^{-\frac{i}{2}k_{\nu }\left\{ x^{\nu },\Im
^{-1}\right\} _{+}}  \label{3.17}
\end{equation}%
a spinor field is written%
\begin{equation}
\Psi =\int d^{4}k\delta (k^{2}-m^{2})\left\{ b_{k}u_{k}e^{-\frac{i}{2}k_{\nu
}\left\{ x^{\nu },\Im ^{-1}\right\} _{+}}+d_{k}^{\ast }v_{k}e^{\frac{i}{2}%
k_{\nu }\left\{ x^{\nu },\Im ^{-1}\right\} _{+}}\right\}   \label{3.18}
\end{equation}%
\begin{equation}
\Psi \in U_{\Re }:D\Psi -m\Psi =0  \label{3.19}
\end{equation}%
From the field a wave function is constructed operating on a vacuum state%
\begin{equation}
\psi =\Psi \left\vert 0\right\rangle   \label{3.19a}
\end{equation}%
Notice that both $b_{k},d_{k}^{\ast }$ and the elements of $U_{\Re }$
operate on $\left\vert 0\right\rangle $, in particular $p_{\mu }\left\vert
0\right\rangle =0$. Now, for a massless field, the (extended) Dirac equation
becomes 
\begin{equation}
D\psi =i\gamma ^{a}\partial _{a}\psi =\left( i\gamma ^{\mu }\partial _{\mu
}+i\gamma ^{4}\partial _{4}\right) \psi =0  \label{3.20}
\end{equation}%
Write%
\begin{equation*}
\psi =e^{-\frac{i}{2}k_{\nu }\left\{ x^{\nu },\Im ^{-1}\right\} _{+}}u\left(
k\right) 
\end{equation*}%
From%
\begin{equation}
\begin{array}{lll}
\partial _{\mu }e^{-\frac{i}{2}k_{\nu }\left\{ x^{\nu },\Im ^{-1}\right\}
_{+}} & = & -ik_{\mu }e^{-\frac{i}{2}k_{\nu }\left\{ x^{\nu },\Im
^{-1}\right\} _{+}} \\ 
\partial _{4}e^{-\frac{i}{2}k_{\nu }\left\{ x^{\nu },\Im ^{-1}\right\} _{+}}
& = & -i\epsilon _{5}\ell \left( -k^{\mu }p_{\mu }+\frac{1}{2}k^{2}\right)
e^{-\frac{i}{2}k_{\nu }\left\{ x^{\nu },\Im ^{-1}\right\} _{+}}%
\end{array}
\label{3.21}
\end{equation}%
one obtains, using (\ref{3.21}), (\ref{3.17}) and (\ref{3.19a})%
\begin{equation}
\begin{array}{ccc}
\left( \gamma ^{\mu }k_{\mu }-\gamma ^{5}\ell \frac{1}{2}k^{2}\right)
u\left( k\right) =0 &  & \epsilon _{5}=+1 \\ 
\left( \gamma ^{\mu }k_{\mu }+i\gamma ^{5}\ell \frac{1}{2}k^{2}\right)
u\left( k\right) =0 &  & \epsilon _{5}=-1%
\end{array}
\label{3.22}
\end{equation}%
Let $\epsilon _{5}=-1$. Iterating (\ref{3.22})%
\begin{equation}
\left( k^{2}-\frac{\ell ^{2}}{4}\left( k^{2}\right) ^{2}\right) u\left(
k\right) =0  \label{3.23}
\end{equation}%
This equation has two solutions, the massless solution ($k^{2}=0$) and
another one, of large mass ($\ell $ being small)%
\begin{equation}
k^{2}=\frac{4}{\ell ^{2}}  \label{3.24}
\end{equation}%
For $\epsilon _{5}=+1$ one would obtain $k^{2}=0$ and%
\begin{equation}
k^{2}=-\frac{4}{\ell ^{2}}  \label{3.24a}
\end{equation}%
a tachyonic large $\left\vert k^{2}\right\vert $ solution.

The solutions of the extended Dirac equation for $k^{2}=0$ are the usual
ones. To find the nature of the solutions for $\left\vert k^{2}\right\vert =%
\frac{4}{\ell ^{2}}$, $\epsilon _{5}=-1$ and $+1$, use a Majorana imaginary
representation for the gamma matrices%
\begin{eqnarray}
\gamma ^{0} &=&\left( 
\begin{array}{cc}
0 & \sigma _{2} \\ 
\sigma _{2} & 0%
\end{array}%
\right) ;\gamma ^{1}=\left( 
\begin{array}{cc}
i\sigma _{1} & 0 \\ 
0 & i\sigma _{1}%
\end{array}%
\right) ;\gamma ^{2}=\left( 
\begin{array}{cc}
0 & \sigma _{2} \\ 
-\sigma _{2} & 0%
\end{array}%
\right)  \notag \\
\gamma ^{3} &=&\left( 
\begin{array}{cc}
i\sigma _{3} & 0 \\ 
0 & i\sigma _{3}%
\end{array}%
\right) ;\gamma ^{5}=i\gamma ^{0}\gamma ^{1}\gamma ^{2}\gamma ^{3}=\left( 
\begin{array}{cc}
-\sigma _{2} & 0 \\ 
0 & \sigma _{2}%
\end{array}%
\right)  \label{3.25}
\end{eqnarray}

\subsection{$\protect\epsilon _{5}=-1,k^{2}=\frac{4}{\ell ^{2}}$}

In the rest frame $k=(m_{0}=\pm \frac{2}{\ell },0,0,0)$. The second equation
in (\ref{3.22}) leads to%
\begin{equation*}
\left( \pm \gamma ^{0}+i\gamma ^{5}\right) u=0
\end{equation*}%
\begin{equation*}
\begin{array}{ccc}
u=\left( 
\begin{array}{c}
a \\ 
ia%
\end{array}%
\right)  &  & \text{Positive energy }\left( m_{0}=\frac{2}{\ell }\right)  \\ 
u=\left( 
\begin{array}{c}
a \\ 
-ia%
\end{array}%
\right)  &  & \text{Negative energy }\left( m_{0}=-\frac{2}{\ell }\right) 
\end{array}%
\end{equation*}%
where $a$ is an arbitrary two-vector. The solutions of non-zero momentum are
obtained by the application of a proper Lorentz transformation $\exp \left( i%
\frac{1}{2}\alpha _{\mu \nu }\left\{ \gamma ^{\mu },\gamma ^{\nu }\right\}
_{+}\right) $.

One has $u^{\ast }\neq u$, hence these solutions are Dirac spinors.

\subsection{$\protect\epsilon _{5}=+1,k^{2}=-\frac{4}{\ell ^{2}}$}

Here one makes $k=(0,0,0,\frac{2}{\ell })$, obtaining%
\begin{equation*}
\left( \gamma ^{3}-\gamma ^{5}\right) u^{\prime }=0
\end{equation*}%
Making $u^{\prime }=$ $\left( 
\begin{array}{c}
a \\ 
b%
\end{array}%
\right) $, $a$ and $b$ being two-vectors, yields%
\begin{eqnarray*}
\left( \sigma _{2}+i\sigma _{3}\right) a &=&0 \\
\left( \sigma _{2}-i\sigma _{3}\right) b &=&0
\end{eqnarray*}%
meaning that $a$ and $b$ are independent two vectors%
\begin{equation*}
a=\left( 
\begin{array}{c}
a_{1} \\ 
a_{1}%
\end{array}%
\right) ;b=\left( 
\begin{array}{c}
b_{1} \\ 
-b_{1}%
\end{array}%
\right) 
\end{equation*}%
with $a_{1}$ and $b_{1}$ real numbers $u^{\prime \ast }=u$ and this
tachyonic, large $\left\vert k^{2}\right\vert $, solution is a Majorana
spinor.

As before, general solutions would be obtained by the action of a Lorentz
transformation.\ 

\section{Coupling the two solutions}

In the previous section it was seen how the extended Dirac equation,
following from the exterior algebra of noncommutative space-time, has both a
massless and a large $\left\vert k^{2}\right\vert $ solution, large if $\ell 
$ is small. If, for example, $\ell $ is in the $10^{-27}-10^{-26}$ seconds
range, $M=\frac{2}{\ell }$ would be of the order of $1$TeV. If the two
solutions mix, one expects that the massless solution would acquire a small
mass as in the seesaw mechanism proposed for neutrinos. In the seesaw
mechanism the large mass (of right-handed neutrinos) is hypothesized to be
obtained from the lepton number violation scale at grand unification. Here
the large mass arises from an independent solution of the same extended
equation.

Let us call $u_{1}$ the zero mass solution and $u_{2}$ the large mass
solution. Then let, us assume that they are coupled by interaction with a
background scalar field that acquires a nonzero vacuum expectation value $%
\phi =\left\langle \phi \right\rangle +h$, with Lagrangian%
\begin{equation}
\mathcal{L}=\overline{u_{1}}i\gamma ^{\mu }\partial _{\mu }u_{1}+\overline{%
u_{2}}\left( i\gamma ^{\mu }\partial _{\mu }+\gamma ^{4}\frac{2}{\ell }%
\right) u_{2}+\left( g\overline{u_{1}}\left( \left\langle \phi \right\rangle
+h\right) u_{2}+h.c.\right)   \label{4.1}
\end{equation}%
leading to the equations of motion:%
\begin{equation}
\begin{array}{ccc}
i\gamma ^{\mu }\partial _{\mu }u_{1}+g\left( \left\langle \phi \right\rangle
+h\right) u_{2} & = & 0 \\ 
\left( i\gamma ^{\mu }\partial _{\mu }+\gamma ^{4}\frac{2}{\ell }\right)
u_{2}+g^{\ast }\left( \left\langle \phi \right\rangle +h^{\ast }\right) u_{1}
& = & 0%
\end{array}
\label{4.2}
\end{equation}%
With $\frac{2}{\ell }$ large, one neglects the kinetic term in the last
equation and obtains, in leading order,%
\begin{equation}
\begin{array}{ccc}
u_{2}\simeq -\frac{\ell }{2}g^{\ast }\left\langle \phi \right\rangle \gamma
^{5}u_{1} &  & \epsilon _{5}=+1 \\ 
u_{2}\simeq i\frac{\ell }{2}g^{\ast }\left\langle \phi \right\rangle \gamma
^{5}u_{1} &  & \epsilon _{5}=-1%
\end{array}
\label{4.3}
\end{equation}%
Substitution in the first equation of (\ref{4.2}) yields

\subsection{$\protect\epsilon _{5}=-1$}

\begin{equation*}
\left( i\gamma ^{\mu }\partial _{\mu }+i\left\vert g\right\vert
^{2}\left\langle \phi \right\rangle ^{2}\frac{\ell }{2}\gamma ^{5}\right)
u_{1}\simeq 0
\end{equation*}%
which has a small mass solution with%
\begin{equation*}
k^{2}=\left( \left\vert g\right\vert ^{2}\left\langle \phi \right\rangle ^{2}%
\frac{\ell }{2}\right) ^{2}
\end{equation*}%
For $k=\left( \pm \left\vert g\right\vert ^{2}\left\langle \phi
\right\rangle ^{2}\frac{\ell }{2},0,0,0\right) $ the solutions are%
\begin{equation*}
u_{1}=\left( 
\begin{array}{c}
a \\ 
\pm ia%
\end{array}%
\right) 
\end{equation*}%
a small mass Dirac spinor. $a$ is an arbitrary two-vector.

\subsection{$\protect\epsilon _{5}=+1$}

\begin{equation*}
\left( i\gamma ^{\mu }\partial _{\mu }-\left\vert g\right\vert
^{2}\left\langle \phi \right\rangle ^{2}\frac{\ell }{2}\gamma ^{5}\right)
u_{1}^{\prime }\simeq 0
\end{equation*}%
which has a small $\left\vert k^{2}\right\vert $ solution with%
\begin{equation*}
k^{2}=-\left( \left\vert g\right\vert ^{2}\left\langle \phi \right\rangle
^{2}\frac{\ell }{2}\right) ^{2}
\end{equation*}%
For $k=\left( 0,0,0,\left\vert g\right\vert ^{2}\left\langle \phi
\right\rangle ^{2}\frac{\ell }{2}\right) $ the solution is%
\begin{equation*}
u_{1}^{\prime }=\left( 
\begin{array}{c}
a_{1} \\ 
a_{1} \\ 
b_{1} \\ 
-b_{1}%
\end{array}%
\right) 
\end{equation*}%
a small $\left\vert k^{2}\right\vert $ tachyonic Majorana spinor.

\end{document}